# Heavy tails and pruning in programmable photonic circuits

Sunkyu Yu[1†] and Namkyoo Park[2*]

[1]Intelligent Wave Systems Laboratory, Department of Electrical and Computer Engineering, Seoul National University, Seoul 08826, Korea

[2]Photonic Systems Laboratory, Department of Electrical and Computer Engineering, Seoul National University, Seoul 08826, Korea

E-mail address for correspondence: [†]sunkyu.yu@snu.ac.kr, [*]nkpark@snu.ac.kr

**Abstract**

Developing hardware for high-dimensional unitary operators plays a vital role in implementing quantum computations and deep learning accelerations. Programmable photonic circuits are singularly promising candidates for universal unitaries owing to intrinsic unitarity, ultrafast tunability, and energy efficiency of photonic platforms. Nonetheless, when the scale of a photonic circuit increases, the effects of noise on the fidelity of quantum operators and deep learning weight matrices become more severe. Here we demonstrate a nontrivial stochastic nature of large-scale programmable photonic circuits—heavy-tailed distributions of rotation operators—that enables the development of high-fidelity universal unitaries through designed pruning of superfluous rotations. The power law and the Pareto principle for the conventional architecture of programmable photonic circuits are revealed with the presence of hub phase shifters, allowing for the application of network pruning to the design of photonic hardware. We extract a universal architecture for pruning random unitary matrices and prove that "the bad is sometimes better to be removed" to achieve high fidelity and energy efficiency. This result lowers the hurdle for high fidelity in large-scale quantum computing and photonic deep learning accelerators.

# Introduction

A unitary operation is an essential building block of quantum[1-3] and classical[4,5] linear systems because any linear operator can be decomposed into a set of unitary and diagonal operators[6]. With advances in quantum computations[3] and deep learning accelerators[7], development of reconfigurable hardware for universal unitary operations has become a topic of intense study. A programmable photonic circuit is one of the most widely used platforms[8,9] for unitary operations in optical neural networks[4,10], modal decoding[5], and quantum computations[1-3].

The fundamental strategy to realize universal unitary operators is to factorize the target operator of the $n$-degree unitary group U($n$) into the diagonal operators and unitary operators of the lower-degree group, such as SU(2)[11]. These subsystems can be realized with conventional optical elements, such as beam splitters, Mach–Zehnder interferometers (MZIs), and phase shifters, which constitute a programmable photonic circuit with reconfigurable modulation. Although the mesh composed of these unit elements can perform universal unitary operations, the connectivity inside the mesh is nonunique and involves an optimal design issue for more compact and robust platforms[12-15]. To improve the original proposal—the Reck design[11]—for the mesh topology, recent approaches have successfully demonstrated advanced arrangements of two-channel subsystems—the Clements design[12]—and the advantages of utilizing multichannel building blocks—the Saygin design[13].

When each channel of the mesh is assigned as a node, a photonic circuit can be interpreted as a graph network[16] regardless of design strategy. Accordingly, it is logical to seek inspiration from network science[17] to understand and improve the large-scale mesh topology of high-degree unitary groups, which should inherit intriguing features of complex networks. In this context, one promising issue is the degree distribution describing the differentiated importance of network nodes, which has been a hot topic through the concepts of heavy-tailed distributions, hub nodes, and scale-freeness[17-21]. When the multiple decomposition processes are applied to U($n$)[11,12], a natural question arises: Do every decomposition and corresponding optical element contribute equally to the designed unitary operation? The answer to this question is of fundamental and practical importance in quantum physics and photonics for devising more advanced hardware architecture applicable to universal quantum evolutions and deep learning accelerators, especially with large-scale photonic circuits.

In this paper, we reveal that some subsystems are more important than others in the common architecture of large-scale programmable photonic circuits. By applying various statistical models to programmable photonic circuits targeting universal unitaries, we verify that a type of unit rotation operator has a heavy-tailed distribution. This finding shows the presence of hub optical elements and the Pareto principle in photonic circuits, which enables the development of the pruning technique[22] for linear quantum or classical hardware. We demonstrate that the suggested hardware pruning for random unitaries allows for improved fidelity when the elements with noise above a specific threshold are removed. This result provides a novel design strategy for high fidelity and energy efficiency in large-scale quantum computations and photonic deep learning accelerators.

## Results

**Rotation operators in programmable photonic circuits**

Before applying the statistical analysis to large-scale programmable photonic circuits, we revisit the Clements design[12], which is one of the most widely used architectures for universal unitaries. Figure 1 shows a schematic of the photonic circuit for the $n \times n$ unitary matrix $U_n \in$ U($n$) obtained from the Clements design. Both the Reck and Clements designs employ nulling the off-diagonal elements of $U_n$ by sequentially multiplying the programmable unit operations $T_m^l \in U(n)$ ($1 \leq m \leq n-1$, $1 \leq l \leq n$, $m$ and $l$ are integers). $T_m^l$ leads to the SU(2) operation on the Bloch sphere defined for the $m^{th}$ and $(m+1)^{th}$ channels to set the off-diagonal element $(l, m)$ or $(m+1, l)$ to be zero.

To maximally cover the SU(2) group with $T_m^l$, reconfigurable and independent control of the amplitude and phase differences between the $m^{th}$ and $(m+1)^{th}$ channels is necessary[8]. One of the most popular platforms for $T_m^l$ is to utilize two pairs of a stationary MZI and a tunable phase shifter in one arm[8,9,12], which involves two adjustable parameters of $\theta \in [0, \pi/2]$ and $\varphi \in [0, 2\pi)$ (Fig. 1a). While the phase shifts $\theta$ and $\varphi$ correspond to tunable $z$-axis rotations on the Bloch sphere, the stationary MZIs constitute the $-\pi/2$ $x$-axis rotations (Fig. 1b,c). The unit operator then becomes $T_m^l(\theta,\varphi) = R_x^m(-\pi/2)R_z^m(-2\theta)R_x^m(-\pi/2)R_z^m(-\varphi)$, where $R_a^m(\xi)$ is the $\xi$-rotation to the $a$-axis on the $m$-$(m+1)$ Bloch sphere, and $\theta$ and $\varphi$ are determined to satisfy nulling of the $(l, m)$ or $(m+1, l)$ element. The target unitary operator $U_n$ is reproduced with multiple $T_m^l$ operators and the remaining diagonal matrix $D_n$ after nulling, as follows:

$$U_n = D_n \left[ \prod_{\{m,l\} \in S_n} T_m^l(\theta_{m,l}, \varphi_{m,l}) \right], \quad (1)$$

where $S_n$ is the ordered sequence of $\{m, l\}$ pairs determined by the nulling process[12] and $D_n$ is realized with phase shifters (Fig. 1d). By newly defining $S_n$, the Clements design employs the highly symmetric arrangement of the MZIs (Fig. 1e), which decreases the device footprint by half and enhances robustness to optical losses compared to the Reck design (see Supplementary Note S1 for the detailed processes).

The reconfigurability for universal unitary operators is thus realized with the $z$-axis rotation $R_z$ obtained from tunable phase shifts $\theta$ and $\varphi$. As programmable devices, the noise and power consumption of photonic circuits are determined by the performance of modulating optical refractive indices $\Delta n$ in the phase shifters and the following changes of $\theta$ and $\varphi$, as $\sim L\Delta n$, where $L$ is the modulation length. Therefore, the statistical analysis of the two adjustable phases $\theta$ and $\varphi$ is critical in examining the performance of large-scale programmable photonic circuits.

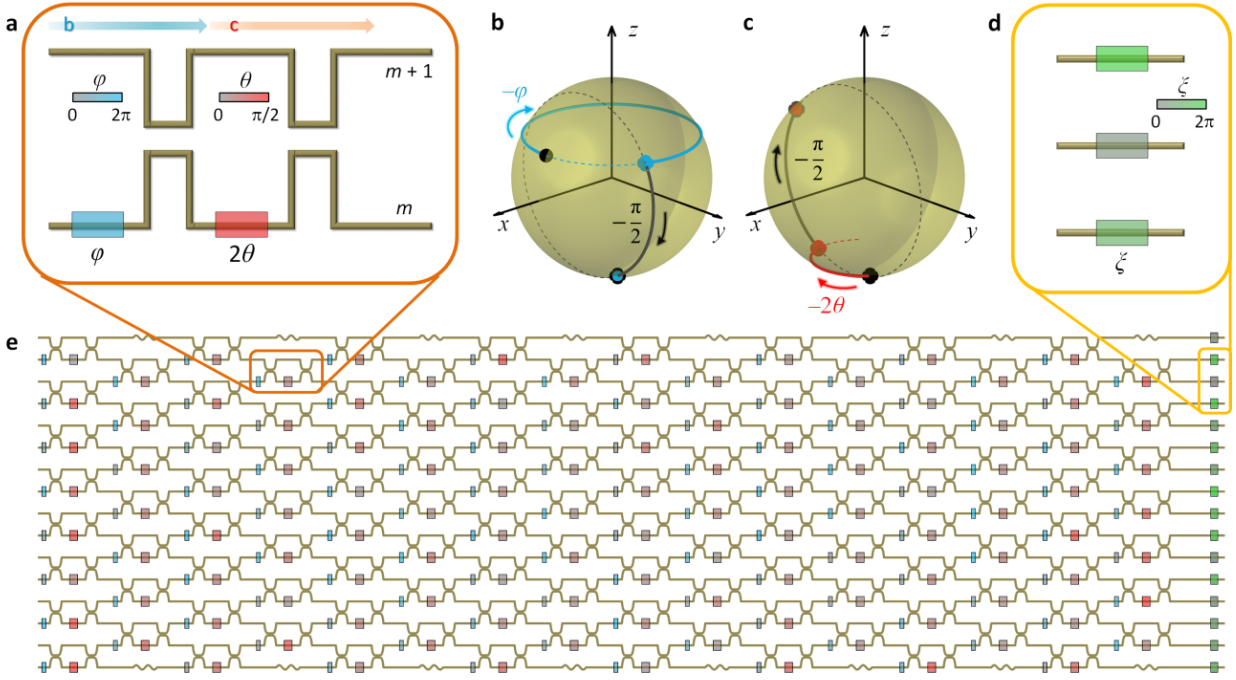

**Fig. 1. Programmable photonic circuits for universal unitary operators. a,** Programmable photonic building block of $T_m^l$ composed of MZIs and phase shifters for the SU(2) operation between the $m^{th}$ and $(m+1)^{th}$ channels. Red and blue boxes represent the phase shifters for $\theta$ and $\varphi$, respectively. **b,c,** The rotation operators of **b,** $R_x^m(-\pi/2)R_z^m(-\varphi)$ and **c,** $R_x^m(-\pi/2)R_z^m(-2\theta)$, described in Bloch spheres. Black and coloured solid lines indicate $x$-axis and $z$-axis rotations, respectively. **b** and **c** correspond to the parts indicated by blue and red arrows in **a**, respectively. **d,** Phase shifters for the diagonal components of $D_n$. **e,** Schematic diagram of the programmable photonic circuit for $U_{16}$. The tunability of $\theta$ and $\varphi$ allows for the programming of $U_{16}$.

**Heavy tails in rotations**

Due to the highly symmetric form of the photonic circuit (Fig. 1e), at first glance, it may appear to be reasonable to predict that the building blocks $T_m^l$ in the circuit have equal importance. Under this presumption, the distributions of $\theta$ and $\varphi$ should be statistically uniform for an ensemble of photonic circuits that generate random unitary operations uniformly distributed in U($n$)[23]. Furthermore, it may also seem reasonable to expect similar distributions for $\theta$ and $\varphi$, both of which perform *z*-axis rotations.

However, upon closer inspection, we reveal that those presumptions are invalid. Instead, there are differences in the contributions of individual building blocks as well as the rotation operators $\theta$ and $\varphi$. First, revisiting the nulling process of the Clements design[12], we note that each off-diagonal element of $U_n$ undergoes differentiated transformations. For example, in nulling the 5 × 5 unitary matrices (Fig. 2a), nulling the (5,1) and (4,1) components results in the $(T_1^5)^\dagger$-transformed 1$^{st}$ and 2$^{nd}$ columns and the $T_3^1$-transformed 3$^{rd}$ and 4$^{th}$ rows, respectively. Because the nulled off-diagonal elements no longer change, each building block treats a matrix element that undergoes a different number of SU(2) transformations; elements that are nulled earlier get fewer transformations (see extended discussion in Supplementary Note S1).

These disparate transformations of each matrix element do not guarantee nontrivial distributions of the phase shifts $\theta$ and $\varphi$. However, the decomposed form of the building block operation $T_m^l(\theta, \varphi) = R_x^m(-\pi/2)R_z^m(-2\theta)R_x^m(-\pi/2)R_z^m(-\varphi)$ results in the nontrivial distribution of $\theta$, which is clearly distinct from that of $\varphi$. Figures 2b and 2c show the transformations of the initial states uniformly distributed in the polar ($\xi$) and azimuthal ($\eta$) axes of the Bloch sphere by multiplying $T_m^l(\theta, \varphi=0) = R_x^m(-\pi/2)R_z^m(-2\theta)R_x^m(-\pi/2)$ and $T_m^l(\theta=0, \varphi) = R_x^m(-\pi)R_z^m(-\varphi)$, respectively, where nonzero $\theta$ and $\varphi$ also have uniformly distributed values in their ranges. Notably, the transformed states by $T_m^l(\theta, \varphi=0)$ become nonuniform (Fig. 2b), in sharp contrast to the uniform distribution from $T_m^l(\theta=0, \varphi)$ (Fig. 2c). Such a discrepancy originates from the difference between the pure *z*-axis rotation $R_z^m(-\varphi)$ and the transformed rotation $R_x^m(-\pi/2)R_z^m(-2\theta)R_x^m(-\pi/2)$ and eventually leads to nonuniformity on the Bloch sphere for $T_m(\theta, \varphi)$ (Fig. 2d). We emphasize that the unequal contributions of each nulling (Fig. 2a) will accumulate the nonuniform distribution from the $\theta$ rotations, which leads to nontrivial statistics in the phase shift design.

To confirm this prediction, we investigate the statistics of $\theta$ and $\varphi$ in realizing programmable photonic circuits that reproduce random unitary operations achieved by uniformly sampling the U($n$) group with the Haar measure[23]. We calculate the probability density functions (PDFs) $p(\theta)$ and $p(\varphi)$ and the complementary cumulative distribution functions (CCDFs) $P(\theta)$ and $P(\varphi)$ for an ensemble of 100 $U_n$ realizations at each $n$. As expected from the uniform distribution with $T_m^l(\theta=0, \varphi)$ (Fig. 2c), the distribution of $p(\varphi)$ is trivially uniform (Supplementary Note S2).

One of the key findings of this work is the nonuniform distribution of $\theta$. Figure 2e shows an example of the $\theta$ distribution for $U_{128}$, which includes 8128 values. As shown in the linearized plots of the CCDF and PDF on the log-log scale, $\theta$ possesses a heavy-tailed distribution[17,19,21], indicating a smaller decrease in $p(\theta)$ for increasing $\theta$ than that of the exponential distribution. For the quantitative analysis, we employ three representative heavy-tailed distribution models[19]—power-law, power-law with an exponential cutoff, and log-normal distributions—and the exponential distribution model. The models are fitted with the $\theta$ data set of each realization of photonic circuits by utilizing analytical or numerical maximization of the model likelihoods[19,24] and the Kolmogorov–Smirnov test[25] for the models with lower bounds (see Methods for details). Notably, all the heavy-tailed models provide good fits for large $n$, showing the consistent behaviours of their estimators for each realization, which is a critical condition for model consistency[21]. For example, the exponent $\alpha$ (Fig. 2f) and the lower bound $\theta_{min}$ (Fig. 2g) in the power-law model $P(\theta) = (\theta/\theta_{min})^{-\alpha+1}$ converge with increasing $n$, which demonstrates that the heavy-tailed distribution becomes more apparent in larger-scale programmable photonic circuits.

The average of the power-law exponents at $n = 128$ is $\alpha_{avg} = 3.18$ for 100 realizations (or 812,800 values of $\theta$), while lower and upper limits are $\alpha_{min} = 2.75$ and $\alpha_{max} = 3.78$. Such consistency clearly proves the validity of the power-law model[20,21] for describing the distribution of the $\theta$-rotation operators (Supplementary Notes S3-S5 for the results of the crossover heavy-tailed distributions and exponential distribution). We note that the averaged lower bound $\theta_{min} = 0.08\pi$ with $P(\theta_{min}) = 0.24$ shows that most of the significant rotations $0.08\pi \leq \theta \leq 0.50\pi$ come from ~24% of the building blocks, which illustrates the Pareto principle for large-scale programmable photonic circuits.

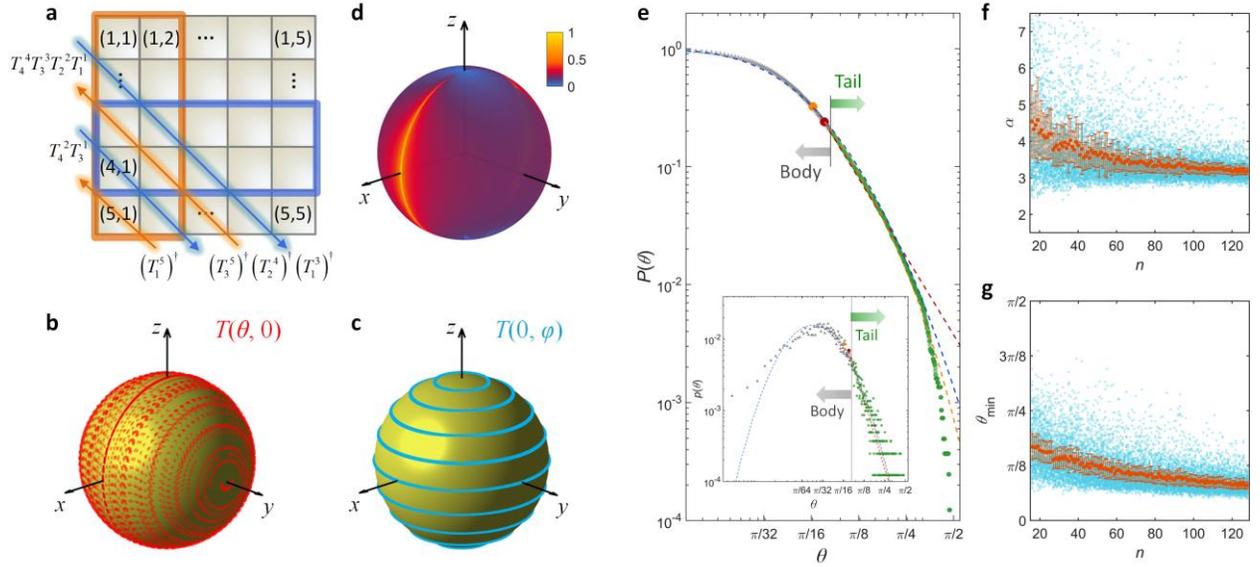

**Fig. 2. Heavy-tailed distributions in unitary photonic circuits. a,b,** Two origins of the heavy-tailed distributions of the rotation operators: **a,** unequal transformations in the nulling process and **b-d,** nonuniform SU(2) rotations. **a,** An example of the nulling process for $U_5$. Orange and green arrows denote the nulling of the off-diagonal elements with $UT^\dagger$ and $TU$, respectively. Red and blue boxes indicate the rotating components for the nulling of the (5,1) and (4,1) components, respectively. **b-d,** Rotated states with **b,** $T(\theta, 0)$, **c,** $T(0, \varphi)$, and **d,** $T(\theta, \varphi)$. Each point in **b** and **c** denotes the transformed state through the corresponding $T$ applied to the uniformly random initial states on the Bloch sphere. The colours in the map in **d** depict the nonuniform density of the transformed states on the Bloch sphere. The initial states in **b** and **c** are obtained with 10 polar grids and 20 azimuthal grids (200 points), while 200 polar grids and 400 azimuthal grids (80,000

points) are used in **d**. **e,** Heavy-tailed distributions of $\theta$ described by the CCDF. The inset shows the PDF and its fitting. The body and tail are separated with $P(\theta) = 0.20$, referring to the Pareto principle. Red, orange, and blue dashed lines show fitting with the power-law, power-law with an exponential cutoff, and log-normal distributions, respectively. Red and orange circles indicate the lower limit of the $\theta$-fitting for the power law and power law with an exponential cutoff, respectively. **f,g,** The variations of the power-law estimators for different $n$: **f,** $\alpha$ and **g,** $\theta_{min}$. Each blue point represents a realization, and orange markers and error bars show the average and root-mean-square error (RMSE) of 100 random realizations at each $n$, respectively.

**Hub units and pruning**

The observed heavy-tailed distribution of $\theta$ rotation operators signifies that some building blocks $T_m^l$ are more critical than others. In realizing programmable photonic circuits for universal unitary operations (Fig. 1a), many phase shifters in the "body" of the distribution may be unnecessary because $\theta \sim 0$. On the other hand, the "tail" phase shifters with large $\theta$ values operate as hub units. Because such hub units deliver most of the necessary $\theta$-rotations for realizing $U_n$, we can envisage the application of the pruning technique in computer science[22] to photonic hardware.

Figure 3a shows the concept of pruning for programmable photonic circuits. The entire photonic circuit for $U_n$ includes $n(n-1)/2$ number of SU(2) building blocks and the same number of $\theta$ values. We define the set of sorted $\theta$ values for a given photonic circuit as $\Theta_n = \{\theta_r | 1 \leq r \leq n(n-1)/2$ for integer $r$, and $\theta_p \leq \theta_q$ for $p \leq q\}$, where $\theta_r$ with larger $r$ represents a more important building block. The pruning of less important ones—body elements—for the photonic circuit is then defined by imposing $\theta_r = 0$ for $1 \leq r \leq \sigma$, where the integer $\sigma$ determines the degree of pruning: $\sigma = 0$ for preserving the original circuit and $\sigma = n(n-1)/2$ for entirely removing $\theta$ rotations in the circuit. In the hardware implementation, pruning corresponds to leaving out the phase shifters for $\theta$ and preserving the symmetry in the MZI arms.

The refractive index modulation in the phase shifters is responsible for much of the energy consumption and noise generation in programmable photonic circuits[8,9]. Therefore, pruning of superfluous phase shifters allows for more energy-efficient and noise-tolerant photonic circuits for reconfigurable unitary operations, provided that the circuit after the pruning accurately reproduces unitary operations. To examine the performance of pruning in a practical situation, we prepare three control groups: one group with the pruning of more important building blocks—tail elements—with $\theta_r = 0$ for $n(n-1)/2 - \sigma + 1 \leq r \leq n(n-1)/2$, and two groups with noisy elements. For the noisy elements, we assume random noise from the phase shifter by assigning the noise $\delta_k$ to the $k^{\text{th}}$ original rotation as $\theta_k + \delta_k$, where $\delta_k = u[0,\delta_0]$ represents the uniform random distribution between 0 and $\delta_0$. For a fair comparison, we construct the groups of noisy elements by replacing the body- or tail-pruned elements in the pruning groups with noisy elements.

To characterize the precision of the operation of the circuits with pruning or noises, we define the fidelity that quantifies the metric between the original and defective operators[26] as follows (see Methods for the derivation):

$$F(U_n^D, U_n^O) = \frac{2\text{Re}\left(\text{Tr}\left[\left(U_n^D\right)^\dagger U_n^O\right]\right)}{N + \text{Tr}\left(\left(U_n^D\right)^\dagger U_n^D\right)} \qquad (2)$$

where $U_n^O$ and $U_n^D$ represent the original unitary matrix and its defective (pruned or noisy) one, respectively, and $\text{Tr}(A)$ is the trace of the square matrix $A$. Figure 3c shows the fidelities of each photonic circuit with the pruning or noise as a function of the ratio of defective elements: $2\sigma/n(n-1)$ in the pruning groups. As expected, the fidelity is preserved much better when the body is pruned instead of the tail. More critical results are shown in comparison with the noisy circuits. When the noise amplitude increases, removing a specific ratio of the "body" phase shifters can be better for higher fidelity than the noisy ones, whether the noise is imposed on body or tail elements. Such a ratio, called the pruning threshold, increases with the noise level and scale of photonic circuits (Fig. 3d). This result states that there is a substantial restriction on the noise level in a large-scale programmable photonic circuit. If a phase shifter cannot meet this restriction, then it is better to remove the phase shifter to increase accuracy and decrease energy consumption for reconfigurability.

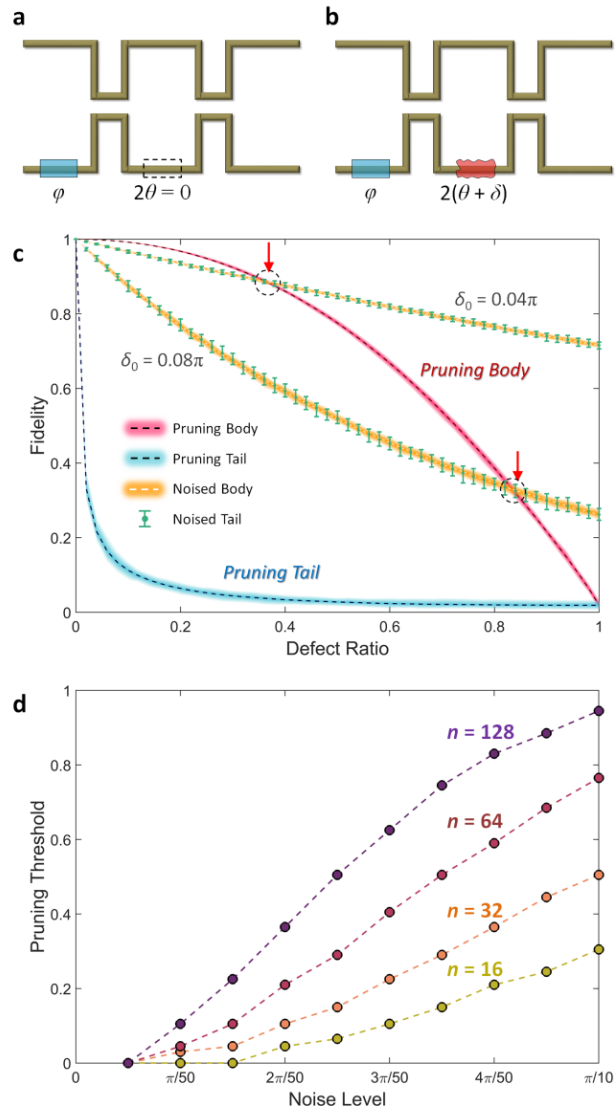

**Fig. 3. Pruning is often better than noise. a,** The concept of pruning in programmable photonic circuits. The phase shifter $2\theta$ of the building block is replaced with an ordinary waveguide, which preserves the symmetry in the MZI arms. **b,** The noisy building block. The phase shifter $2\theta$ is perturbed as $2(\theta + \delta)$. **c,** Comparison of the fidelities of the photonic circuits in different groups: pruning body (red line), pruning tail (blue line), noisy body (orange line), and noisy tail (green error bars). The thicknesses of the coloured lines and the error bars present the range of the fidelities between their maxima and minima. The red arrows indicate the pruning thresholds for each case. Two pairs of groups with noisy bodies and noisy tails are shown for $\delta_0 = 0.04\pi$ and $0.08\pi$. **d,** Pruning threshold as a function of the noise level $\delta_0$ for different degrees of unitary operators. In **c** and **d**, 100 random $U_n$ realizations are analysed per value of $n$ and defect ratio.

**Universal architecture for pruning**

Although the result shown in Fig. 3 demonstrates hub functionality and the advantage of pruning in realizing an individual unitary operator, it is insufficient to apply pruning to programmable photonic circuits for universal unitary operators. This is because the sorted set $\Theta_n$ for pruning varies with the form of a unitary operator. To apply the pruning method for universal unitaries with reconfigurability, it is necessary to construct an adaptable architecture for the pruning process.

Because the position of each building block for nulling a specific off-diagonal element is fixed in an $n$-degree photonic circuit, the averages of the phase rotations $<\theta_{m,l}>$ and $<\varphi_{m,l}>$ are well-defined in hardware for random unitary operations that are uniformly sampled from U($n$). Figures 4a and 4b describe the universal architectures defined by $<\theta_{m,l}>$ and $<\varphi_{m,l}>$, respectively, for 100 $U_n$ realizations of $n = 16$ and $n = 32$ (see Supplementary Note S6 for $n = 64$). As expected from the distinct SU(2) operations from $\theta$ and $\varphi$ (Fig. 2b,c), we observe a spatially inhomogeneous distribution of $<\theta_{m,l}>$ in contrast to that of $<\varphi_{m,l}>$. More specifically, the universal architectures show significant $\theta$-rotation contributions from the building blocks near the boundary of the programmable photonic circuits. Such a consistent distribution allows for a universal sorted set $\langle\Theta\rangle_n = \{\langle\theta_{m,l}\rangle_r | 1 \leq r \leq n(n-1)/2$ for integer $r$ and $\langle\theta_{m,l}\rangle_p \leq \langle\theta_{m,l}\rangle_q$ for $p \leq q\}$ to develop a pruning process applicable to any unitary operations.

From this guideline, we again employ pruning and add noise to the body and tail elements to the set $\langle\Theta\rangle_n$. As shown in Figs 4c and 4d, the general tendencies in Figs 3c and 3d are preserved; the tail is more important than the body, the bad is better to be removed, and pruning is more efficient for larger-scale photonic circuits. Although the minimum noise level increases, there is still a pruning threshold that guarantees the advantage of removing $\theta$ phase shifters, and this tendency is much more apparent in larger-scale programmable photonic circuits. Notably, the importance of protecting hub elements from noise becomes evident at the strong noise level ($\delta_0 = 0.20\pi$ cases in Fig. 4c).

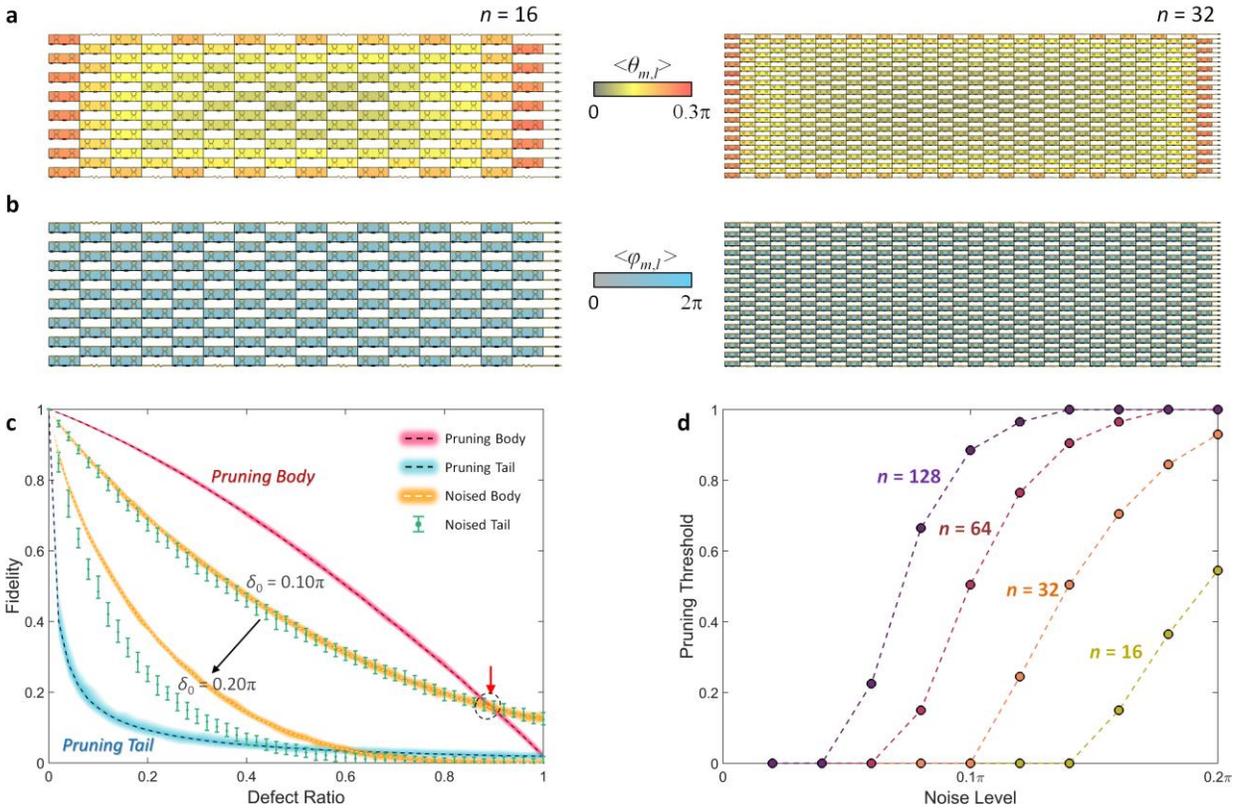

**Fig. 4. Universal architecture for pruning in reconfigurable unitaries. a,b,** The averages of **a,** $\langle\theta_{m,l}\rangle$ and **b,** $\langle\varphi_{m,l}\rangle$ for the photonic circuits of 100 $U_n$ realizations with $n = 16$ and $n = 32$. We set

the upper bound of the colour map in **a** to be 0.3 π for better visibility. **c,** Comparison of the fidelities of the photonic circuits in different groups: pruning body (red line), pruning tail (blue line), noisy body (orange line), and noisy tail (green error bars). The thicknesses of the coloured lines and the error bars present the range of the fidelities between their maxima and minima. The red arrow indicates the pruning threshold. Two pairs of groups with noisy bodies and noisy tails are shown for $\delta_0$ = 0.10π and 0.20π. **d,** Pruning threshold as a function of the noise level $\delta_0$ for different degrees of unitary operators.

## Discussion

Due to the mathematical generality of our study, the presented results should be universal for programmable photonic[8,9] or superconducting[27] processors for reconfigurable unitary operations when the unit SU(2) operation is nonuniform on the Bloch sphere and the target degree *n* is finite. Notably, we observed excellent fitting with the power-law model, the crossover behaviours from exponential to heavy tails in the truncated power-law and log-normal models, and the evident failure of the exponential model nearly above the degree $n \geq 80$. Therefore, the heavy-tailed features are evident at the scale near and beyond the present state-of-the-art degrees ($n \sim 10^2$) in deep learning accelerators[4,28,29] and noisy intermediate-scale quantum (NISQ) computers[30-32]. The suitable application of the demonstrated pruning method, which allows for leaving out a significant portion of electro-optic modulations in programmable photonic circuits, will become particularly beneficial for the next era of quantum computing and deep learning hardware.

The presence of heavy-tailed distributions in programmable photonic circuits inspires the extension of seminal achievements in probability theory and network science to wave physics. As shown in our study, the intriguing features related to heavy-tailed distributions are also demonstrated in wave platforms, such as the observed Pareto principle in wave physics and the critical role of hub elements in pruning and noise immunity. When the connectivity of integrated wave systems becomes more extensive and complex[33], the concepts of complex networks will provide a foundation for novel design strategies in wave physics.

In conclusion, we demonstrated that some of the unit elements in a large-scale programmable photonic circuit are more important than others, exhibiting the heavy-tailed feature verified with conventional statistical models, i.e., the power-law, power-law with an exponential cutoff, and log-normal distributions, and the exponential distribution as a counterexample. The observed heavy-tailed distribution originates from nonuniform rotations on the Bloch sphere, which are ubiquitous in conventional SU(2) units for programmable photonic circuits. The result allows for a new design strategy—pruning—for high fidelity and energy efficiency, which offers intriguing insight into the design of large-scale photonic structures for classical and quantum applications. Further research on devising other forms of SU(2) units or units with higher degree for $U_n$ factorization is desirable to alter the observed heavy-tailed distributions.

## Methods

**Model fitting process.** To analyse the $\theta$ distributions in an ensemble of programmable photonic circuit realizations, we employ multiple statistical models: power-law, power-law with an exponential cutoff, log-normal, and exponential distributions. Each model is defined by a set of model parameters $\{q_s\}$. To calculate the model parameters for the fitting of a given data set $\{\theta_1, \theta_2, …, \theta_M\}$, we employ an analytical or numerical calculation of the maximum likelihood estimators (MLEs)[24]. First, the probability of obtaining the data set from the statistical model with the given model parameters $\{q_s\}$ and the PDF $p(\theta_m;\{q_s\})$ is

$$p(\{\theta_m\};\{q_s\}) = \prod_{m=1}^{M} p(\theta_m;\{q_s\}), \tag{3}$$

which is called the likelihood for the data and model. Because the employed statistical models have exponential forms, it is conventional to utilize the log-likelihood $L$:

$$L = \sum_{m=1}^{M} \log\left(p(\theta_m;\{q_s\})\right). \tag{4}$$

The fitting of the model to a given set of data, which requires the calculation of $\{q_s\}$, then corresponds to the maximization of $L$ with respect to $\{q_s\}$. Therefore, the MLE is defined as

$$\nabla_{\{q_s\}} L = \nabla_{\{q_s\}} \sum_{m=1}^{M} \log\left(p(\theta_m;\{q_s\})\right) = O. \tag{5}$$

**Power-law distribution model.** In analysing the heavy-tailed statistics of the $\theta$-rotations, we mainly employ the power-law distribution model[17-19], which supports the PDF and CCDF, as follows:

$$p(\theta) = \frac{\alpha - 1}{\theta_{\min}} \left(\frac{\theta}{\theta_{\min}}\right)^{-\alpha}, \tag{6}$$

$$P(\theta) = \left(\frac{\theta}{\theta_{\min}}\right)^{-\alpha+1}, \tag{7}$$

where $\alpha$ and $\theta_{\min}$ are the exponent and lower bound of the power-law model, respectively. The model is defined in the range $\alpha > 1$, and the model parameter set is $\{q_s\} = \{\alpha\}$. For a given data set, the log-likelihood becomes

$$L = M \log(\alpha - 1) + M(\alpha - 1) \log \theta_{\min} - \alpha \sum_{m=1}^{M} \log \theta_m. \tag{8}$$

The MLE then leads to $\alpha$, as

$$\alpha = 1 + M \left[\sum_{m=1}^{M} \log\left(\frac{\theta_m}{\theta_{\min}}\right)\right]^{-1}. \tag{9}$$

We calculate an array of $\alpha$ values using Eq. (9) for all the possible values of $\theta_{\min}$, where each pair of $\alpha$ and $\theta_{\min}$ comprises a candidate power-law model.

**Power-law model with an exponential cutoff.** To obtain a thorough confirmation of the heavy-tailed statistics, we test crossover distributions between a power-law and an exponential distribution. First, we apply the power-law model with an exponential cutoff, which is the truncated version of the original power-law model. The PDF and CCDF of the model are[17,19]:

$$p(\theta) = \frac{\lambda_c^{1-\alpha_c}}{\Gamma(1-\alpha_c, \lambda_c \theta_{c,\min})} \theta^{-\alpha_c} e^{-\lambda_c \theta}, \tag{10}$$

$$P(\theta) = \frac{\Gamma(1-\alpha_c, \lambda_c \theta)}{\Gamma(1-\alpha_c, \lambda_c \theta_{c,\min})}, \tag{11}$$

where $\alpha_c$, $\lambda_c$, and $\theta_{c,\min}$ are the power-law exponent, cutoff exponent, and the lower bound of the model, respectively, and $\Gamma(s,x)$ is the upper incomplete gamma function. The model is defined in the range of $\alpha_c \geq 0$ and $\lambda_c \geq 0$. The log-likelihood for the data set $\{\theta_1, \theta_2, \ldots, \theta_M\}$ is

$$L = M(1-\alpha_c)\log \lambda_c - M \log \Gamma(1-\alpha_c, \lambda_c \theta_{c,\min}) - \alpha_c \sum_{m=1}^{M} \log \theta_m - \lambda_c \sum_{m=1}^{M} \theta_m. \tag{12}$$

Although the MLE with the model parameters $\{q_s\} = \{\alpha_c, \lambda_c\}$ leads to the following relations:

$$\log \lambda_c + \frac{\partial_{\alpha_c} \Gamma(1-\alpha_c, \lambda_c \theta_{c,\min})}{\Gamma(1-\alpha_c, \lambda_c \theta_{c,\min})} = -\frac{1}{M}\sum_{m=1}^{M}\log \theta_m,$$

$$\frac{1-\alpha_c}{\lambda_c} + \frac{\theta_{c,\min}(\lambda_c \theta_{c,\min})^{-\alpha_c} e^{-\lambda_c \theta_{c,\min}}}{\Gamma(1-\alpha_c, \lambda_c \theta_{c,\min})} = \frac{1}{M}\sum_{m=1}^{M}\theta_m, \tag{13}$$

we instead employ the numerical minimization of $-L$ with the constraints $\alpha_c \geq 0$ and $\lambda_c \geq 0$ due to the difficulty in handling the analytical derivative of the upper incomplete gamma function. We calculate the pairs of $\alpha_c$ and $\lambda_c$ for all the possible values of $\theta_{c,\min}$, where a set of $\alpha_c$, $\lambda_c$, and $\theta_{c,\min}$ comprises a candidate for the model.

**Log-normal distribution model.** To cover the intermediate regime between the power-law and exponential distributions[17], we employ another crossover distribution: the log-normal distribution model. The PDF and CCDF of the model are[17,19]:

$$p(\theta) = \frac{1}{\sigma\theta\sqrt{2\pi}} \exp\left(-\frac{(\log\theta - \mu)^2}{2\sigma^2}\right), \quad (14)$$

$$P(\theta) = \frac{1}{2}\left[1 - \text{erf}\left(\frac{\log\theta - \mu}{\sigma\sqrt{2}}\right)\right], \quad (15)$$

where $\mu$ and $\sigma$ are the mean and standard deviation of $\log(\theta)$, respectively, and erf is the error function. With the model parameters $\{q_s\} = \{\mu, \sigma\}$, the log-likelihood and the MLE relation are shown in Eqs (16) and (17), respectively, as follows:

$$L = -\sum_{m=1}^{M}\log\theta_m - M\log\sigma - \frac{M}{2}\log 2\pi - \sum_{m=1}^{M}\frac{(\log\theta_m - \mu)^2}{2\sigma^2}. \quad (16)$$

$$\sum_{m=1}^{M}\frac{\log\theta_m - \mu}{\sigma^2} = 0,$$
$$\frac{M}{\sigma} = \sum_{m=1}^{M}\frac{(\log\theta_m - \mu)^2}{\sigma^3}. \quad (17)$$

Instead of utilizing the analytical MLE, we employ numerical minimization of $-L$ with the constraint $\sigma \geq 0$.

**Exponential distribution model.** For the comparison with models other than heavy-tailed distributions, we test the exponential distribution model[17,19,21], which has the following PDF and CCDF:

$$p(\theta) = \lambda_e e^{\lambda_e \theta_{e,\min}} e^{-\lambda_e \theta}, \quad (18)$$

$$P(\theta) = e^{\lambda_e \theta_{e,\min}} e^{-\lambda_e \theta}, \quad (19)$$

where the model parameter is $\{q_s\} = \{\lambda_e\}$. The log-likelihood and the MLE relation are

$$L = \log \lambda_e + M \lambda_e \theta_{e,\min} - \lambda_e \sum_{m=1}^{M} \theta_m, \quad (20)$$

$$\lambda_e = \left[ \sum_{m=1}^{M} \theta_m - M \theta_{e,\min} \right]^{-1}. \quad (21)$$

We calculate an array of $\lambda_e$ values using Eq. (21) for all the possible values of $\theta_{e,\min}$, where each pair of $\lambda_e$ and $\theta_{e,\min}$ comprises a candidate for the model.

**Kolmogorov–Smirnov test.** In the power-law, power-law with an exponential cutoff, and exponential distribution models, we obtain multiple candidates for the models with different values of lower bounds $\theta_{\min}$, $\theta_{c,\min}$, and $\theta_{e,\min}$, respectively. Each candidate of a model supports a distinct range of data for model validity and possesses different values of model parameters $\{q_s\}$. To extract the optimum model among the candidates, we apply the Kolmogorov–Smirnov (KS) test[19,25]. When the CDFs of the data set and the statistical model are $S(\theta)$ and $P(\theta; \theta_{\min}, \{q_s\})$ for the lower bound parameter $\theta_{\min}$, we define the maximum distance $D$ between the data and model distributions as:

$$D = \max_{\theta \geq \theta_{\min}} \left| S(\theta) - P(\theta; \theta_{\min}, \{q_s\}) \right|. \quad (22)$$

We select $\theta_{\min}$ and the corresponding $\{q_s\}$ to minimize $D$, determining the optimum statistical model for each case of the power-law, power-law with an exponential cutoff, and exponential distribution models.

**Fidelity for unitary matrices.** We consider the $n \times n$ unitary matrix $U_n^O$ and its defective one $U_n^D$, which could be nonunitary in general. The cost function or the square of the metric between the matrices is defined by[26]:

$$J_U = \frac{1}{N^2} \sum_{i,j} \left| U_n^O{}_{(i,j)} - U_n^D{}_{(i,j)} \right|^2$$
$$= \frac{1}{N} + \frac{1}{N^2} \text{Tr}\left( \left(U_n^D\right)^\dagger U_n^D - 2\,\text{Re}\left[ \left(U_n^D\right)^\dagger U_n^O \right] \right) \tag{23}$$

where $A_{(i,j)}$ is the $(i,j)$ matrix component and $\text{Tr}(A)$ is the trace of the square matrix $A$. Because $J_U \geq 0$, we obtain the relationship:

$$N + \text{Tr}\left( \left(U_n^D\right)^\dagger U_n^D \right) \geq 2\,\text{Tr}\left( \text{Re}\left[ \left(U_n^D\right)^\dagger U_n^O \right] \right), \tag{24}$$

where equality is achieved with the minimum defect, as $U_n^O = U_n^D$. Because the left side of Eq. (24) is positive, the definition of fidelity is $F(U_n^D, U_n^O)$ in Eq. (2) in the main text.

## Data availability

The data that support the plots and other findings of this study are available from the corresponding author upon request.

## Code availability

All code developed in this work will be made available upon request.

**Acknowledgements**

We acknowledge financial support from the National Research Foundation of Korea (NRF) through the Basic Research Laboratory (No. 2021R1A4A3032027), Young Researcher Program (No. 2021R1C1C1005031), and Global Frontier Program (No. 2014M3A6B3063708), all funded by the Korean government.

**Author contributions**

Both authors conceived the idea, discussed the results, and contributed to the final manuscript.

**Competing interests**

The authors have no conflicts of interest to declare.

**Additional information**

**Correspondence and requests for materials** should be addressed to S.Y. or N.P.

**Supplementary Information for "Heavy tails and pruning in programmable photonic circuits"**

Sunkyu Yu[1†] and Namkyoo Park[2*]

[1]Intelligent Wave Systems Laboratory, Department of Electrical and Computer Engineering, Seoul National University, Seoul 08826, Korea

[2]Photonic Systems Laboratory, Department of Electrical and Computer Engineering, Seoul National University, Seoul 08826, Korea

E-mail address for correspondence: [†]sunkyu.yu@snu.ac.kr, [*]nkpark@snu.ac.kr

**Note S1. Differentiated transformations of off-diagonals through nulling processes**

**Note S2. Uniform distribution of $\varphi$-rotations**

**Note S3. Heavy-tailed distribution: Power-law model with an exponential cutoff**

**Note S4. Heavy-tailed distribution: Log-normal model**

**Note S5. Non-heavy-tailed distribution: Exponential model**

**Note S6. Universal architecture for $n = 64$**

**Note S1. Differentiated transformations of off-diagonals through nulling processes**

Figure S1 describes the first 4 steps of nulling the off-diagonal elements in the 5-degree random unitary matrix $U_5$. Each nulling process is achieved with $T_m^l$, which has the SU(2) block matrix for the $m^{th}$ and $(m+1)^{th}$ channels (green squares in $T_m^l$ or $(T_m^l)^\dagger$). The nulling processes are composed of two forms of nulling to preserve the nulled elements (black squares in $U$ and $U'$) from the prior steps: $U = U'(T_m^l)^\dagger$ (orange arrows in Figs S1a and S1d) and $U = T_m^l U'$ (blue arrows in Figs S1b and S1c), where $U$ and $U'$ are the transformed unitary matrix at the current and previous steps, respectively. Each form sets one of the off-diagonal elements to be zero: $(l, m)$ matrix element with $U'(T_m^l)^\dagger$ and $(m+1, l)$ matrix element with $T_m^l U'$.

During the nulling of the target off-diagonal element with the designed $\theta$ and $\varphi$, the other elements at the nearby rows or columns (orange or blue boxes in $U$ and $U'$) inevitably undergo SU(2) operations. Because the values of the previously nulled off-diagonal elements are maintained, each matrix element eventually undergoes a different number of SU(2) transformations; elements that are nulled earlier get fewer transformations. Such a discrepancy in the numbers of SU(2) transformations applied to the off-diagonal elements is general for any design strategies that adopt a series of nulling processes for the factorization of the target unitary operation, such as the Reck design[1]. However, the specific number of SU(2) transformations of each matrix element can differ depending on the nulling algorithm.

As described in the Clements design[2], nulling processes are applied only to the lower triangular off-diagonal elements because a unitary triangular matrix is diagonal. After the entire nulling process, we employ the relationship $(T_m^l)^\dagger D = D' T_m^l$, where $D$ is the diagonal matrix resulting from the nulling processes, and $D'$ is its transformation with $T_m^l$. This relation that is valid for Hermitian systems is used to derive Eq. (1) in the main text.

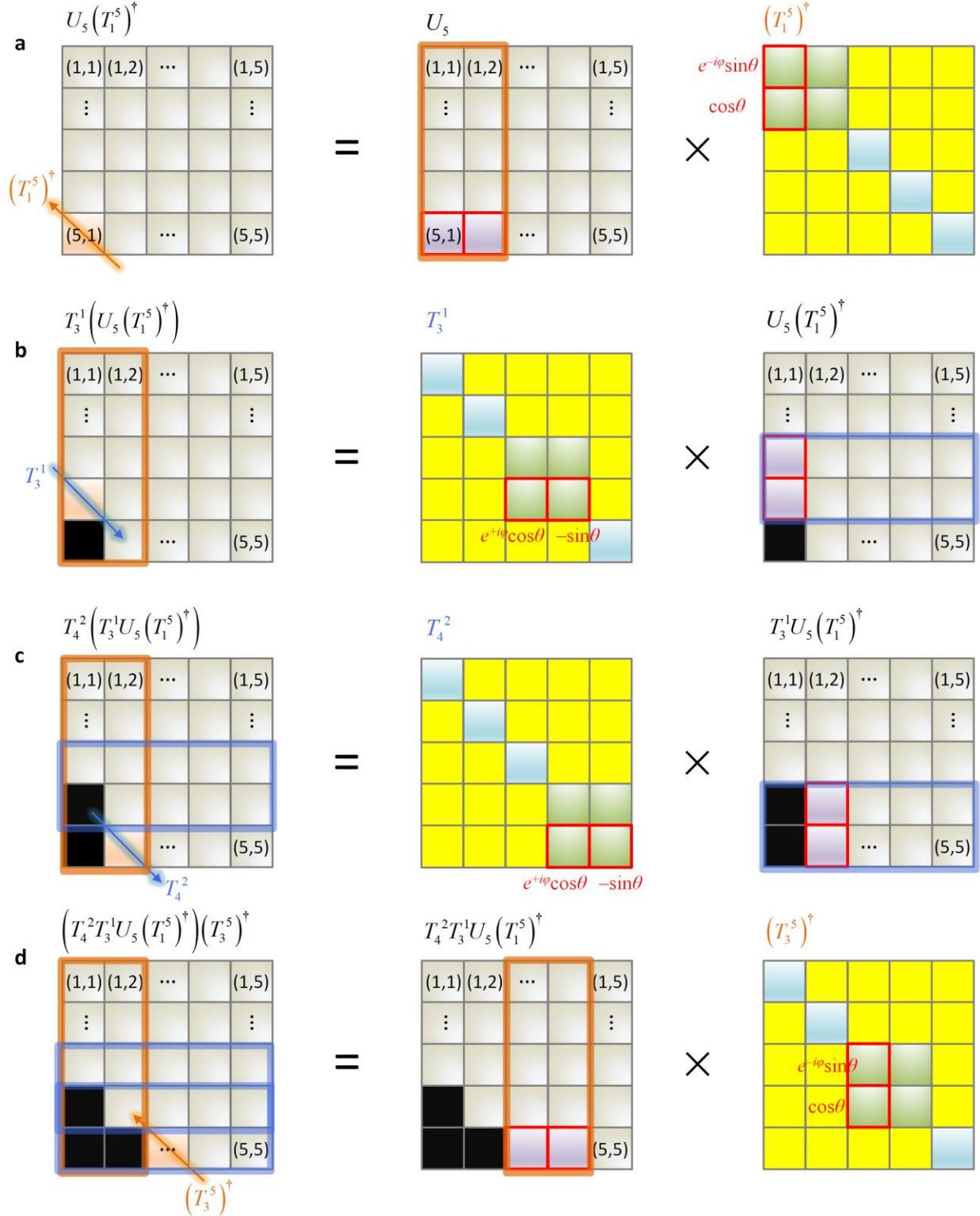

**Fig. S1. Why are nulling processes unequal? a-d,** The first 4 transformations of the 5-degree random unitary matrix $U_5$: **a,** $U = U_5(T_1^5)^\dagger$, **b,** $U = T_3^1 U' = T_3^1 U_5(T_1^5)^\dagger$, **c,** $U = T_4^2 U' = T_4^2 T_3^1 U_5(T_1^5)^\dagger$, and **d,** $U = U'(T_3^5)^\dagger = T_4^2 T_3^1 U_5(T_1^5)^\dagger(T_3^5)^\dagger$. $U$ and $U'$ are the transformed unitary matrices at the
3

current and previous steps of the nulling processes. Orange arrows in **a** and **d** and blue arrows in **b** and **c** represent the nulling of the target off-diagonal element in $U$. Purple and green squares surrounded by red boxes represent the multiplied elements in $U'$ and $T_m^l$ (or $(T_m^l)^\dagger$), respectively, to obtain the target element in $U$. This multiplication for nulling defines the values of $\theta$ and $\varphi$. Black squares in $U$ and $U'$ denote nulled elements. The orange and blue boxes in $U$ and $U'$ are matrix elements that undergo SU(2) transformations from $T_m^l$. Yellow and blue squares in $T_{ml}$ or $(T_m^l)^\dagger$ represent values of 0 and 1, respectively.

**Note S2. Uniform distribution of $\varphi$-rotations**

Figure S2 shows an example of the $\varphi$ distribution for $U_{128}$, which is obtained from the uniform sampling of the U(128) group with the Haar measure[3]. The ensemble includes 8128 values of $\varphi$, which is half the number of off-diagonal elements. As shown in the linearized CCDF plot on the linear scale (Fig. S2a) and the almost flat distribution of the PDF (Fig. S2b), $\varphi$ possesses a uniform distribution in sharp contrast to the heavy-tailed distribution of $\theta$. This result proves that the differentiated behaviours of $\theta$ and $\varphi$ on the Bloch sphere (Figs 2b and 2c in the main text) result in the apparent distinction in their statistics.

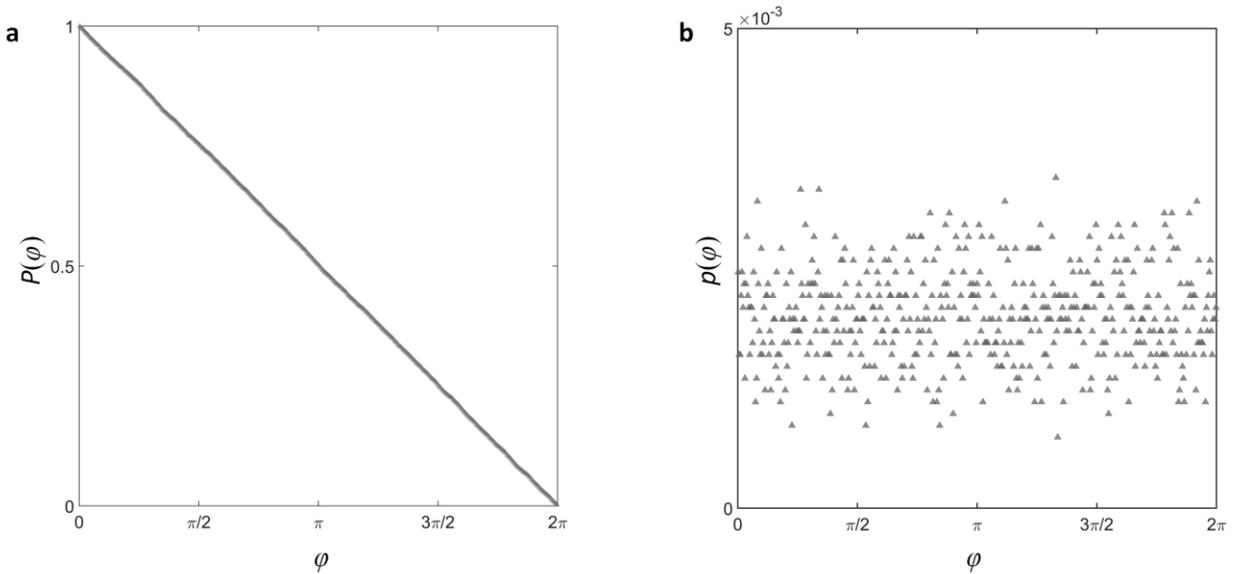

**Fig. S2. Statistical distributions of $\varphi$-rotations. a,b,** Distributions of $\varphi$ described by its **a,** CCDF, and **b,** PDF. The range of $\varphi$ is $\varphi \in [0, 2\pi)$. All the scales of the axes are linear.

**Note S3. Heavy-tailed distribution: Power-law model with an exponential cutoff**

Although the power-law model is one of the well-established heavy-tailed models[4-6], many empirical data sets in natural, technological, or social systems are not perfectly described by the ideal power-law model. To handle imperfect types of power-law-like data sets, some truncated forms of the power-law model, such as the power-law model having a cutoff, can be successful.

We apply the power-law model with an exponential cutoff to the $\theta$-distributions of programmable photonic circuits with different degrees $n$. The model parameters $\{q_s\} = \{\alpha_c, \lambda_c\}$ and the lower-bound parameter $\theta_{c,\min}$ are obtained from the minimization of $-L$ and the Kolmogorov–Smirnov test, as described in the Methods section. Similar to the results for the power-law model (Figs 2f and 2g in the main text), the model parameters and the lower-bound parameter become consistent with the increase in $n$, which shows the validity of the model (Fig. S3). Notably, the model parameters thoroughly describe the crossover in our problem. The increase in the power-law exponent $\alpha_c$ and the decrease in the cutoff exponent $\lambda_c$ (Figs S3a and S3b) demonstrate that the tails of the distribution become heavier with a larger $n$, which is also confirmed by the increasing range of model validity (Fig. S3c).

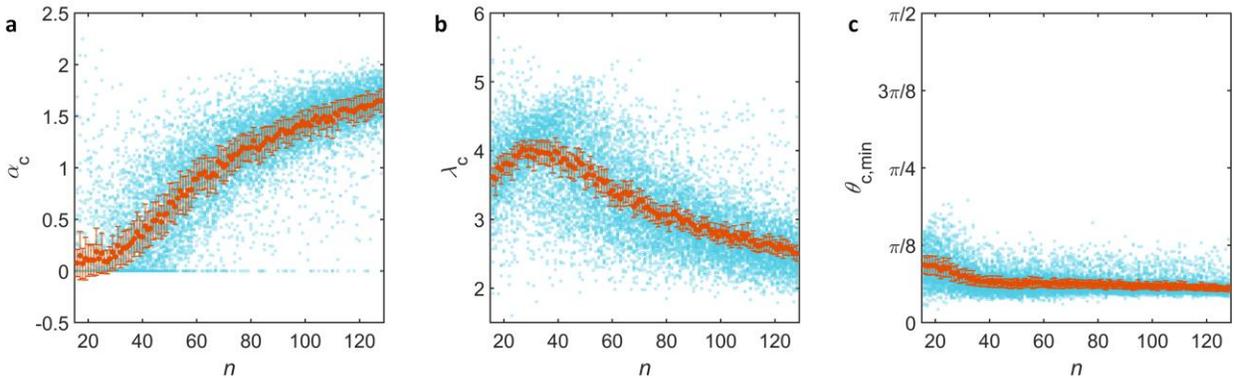

**Fig. S3. Model parameters of the power-law distribution with an exponential cutoff. a-c,** The variations of the model estimators for different $n$: **a,** $\alpha_c$, **b,** $\lambda_c$, and **c,** $\theta_{c,\min}$. Each blue point represents a realization, and orange markers and error bars show the average and root-mean-square error (RMSE) of 100 realizations at each value of $n$, respectively.

**Note S4. Heavy-tailed distribution: Log-normal model**

We also study another form of the heavy-tailed model: the log-normal distribution, which can originate from the product of multiple independent positive random numbers[4]. Figure S4 shows the $n$-dependent model parameters $\{q_s\} = \{\mu, \sigma\}$ obtained from the minimization of $-L$ (Methods section), which demonstrates the model validity for large-scale photonic circuits with the enhanced consistency of $\mu$ and $\sigma$.

As a crossover distribution between a power-law model and an exponential model, the log-normal distribution with large $\sigma$ resembles a power law due to the exponential growth of $<\theta^2>$, where $<...>$ is the ensemble average[4]. In this context, Fig. S4b further demonstrates a more power-law-like (or heavy-tailed) nature of the $\theta$-distribution in larger-scale photonic circuits.

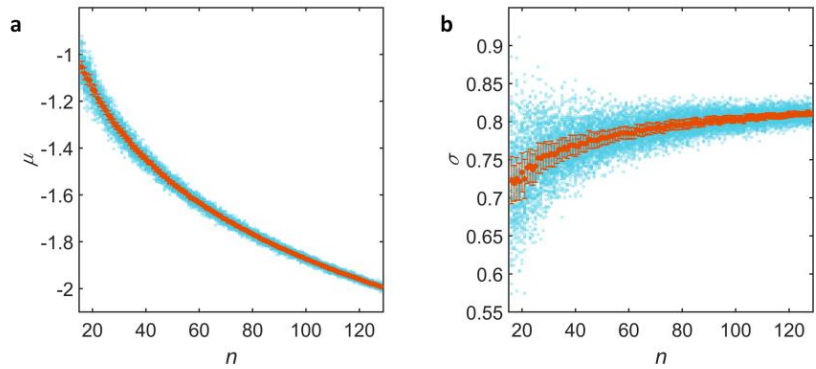

**Fig. S4. Model parameters of the log-normal distribution. a,b,** The variations of the model estimators for different $n$: **a,** $\mu$, and **b,** $\sigma$. Each blue point represents a realization, and orange markers and error bars show the average and root-mean-square error (RMSE) of 100 realizations at each $n$, respectively.

**Note S5. Non-heavy-tailed distribution: Exponential model**

While we demonstrate the validity of multiple heavy-tailed models—power-law, power-law with an exponential cutoff, and log-normal models—for $\theta$-distributions, it is instructive to compare the use of a distribution other than a heavy-tailed model, such as the exponential distribution, to fit the data[4,7]. Figure S5 shows the model parameter $\{q_s\} = \{\lambda_e\}$ and the lower-bound parameter $\theta_{e,min}$ obtained from the MLE relation and the Kolmogorov–Smirnov test (Methods section). As shown in Fig. S5a, the model is inconsistent with the significant deviations from the average of $\lambda_e$. While the level of such errors from the model is maintained regardless of $n$ values, the lower bound $\theta_{e,min}$ increases with $n$ (Fig. S5b), demonstrating that the exponential model does not provide a valid fit to $\theta$-distributions.

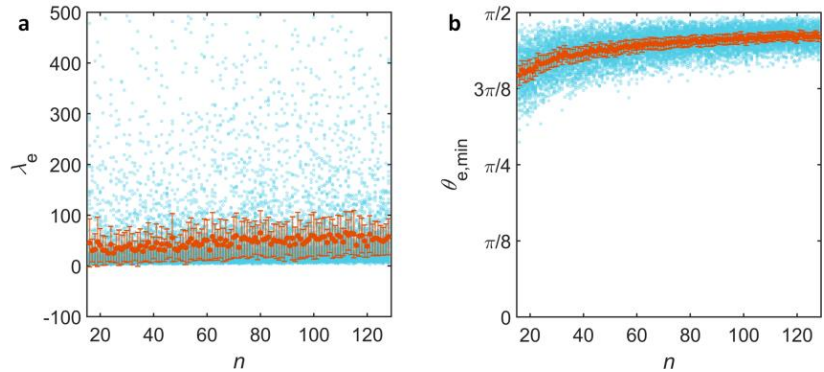

**Fig. S5. Model parameters of the exponential distribution. a,b,** The variations of the model estimators for different $n$: **a,** $\lambda_e$ and **b,** $\theta_{e,min}$. Each blue point represents a realization, and orange markers and error bars show the average and root-mean-square error (RMSE) of 100 realizations at each $n$, respectively.

**Note S6. Universal architecture for *n* = 64**

Figures S6a and S6b describe the universal architectures defined by $\langle\theta_{m,l}\rangle$ and $\langle\varphi_{m,l}\rangle$, respectively, for *n* = 64. Similar to the results of *n* = 16 and *n* = 32 (Figs 4a and 4b in the main text), we again confirm the inhomogeneous distribution of $\langle\theta_{m,l}\rangle$ in contrast to the uniform distribution of $\langle\varphi_{m,l}\rangle$.

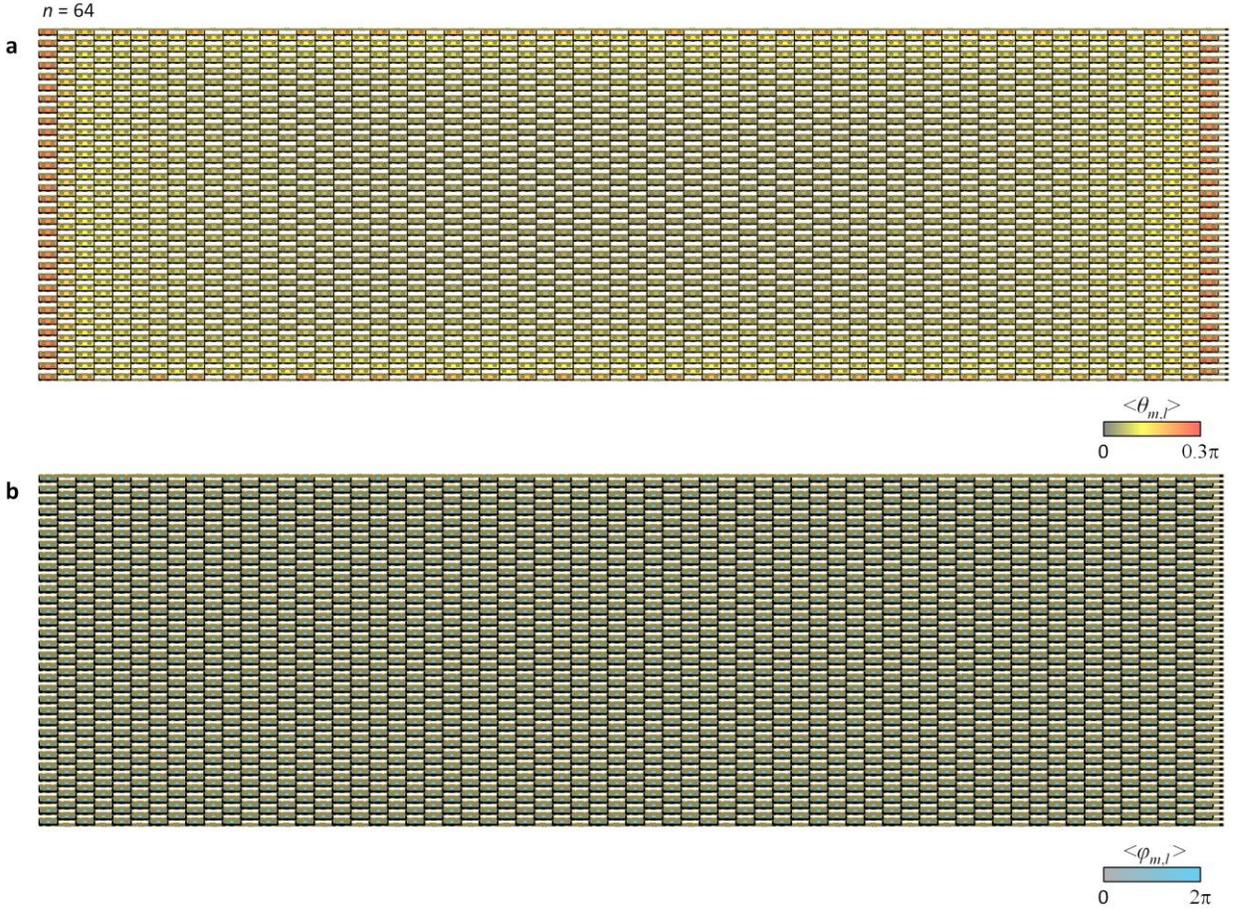

**Fig. S6. Universal architecture of programmable photonic circuits for $U_{64}$. a,b,** The averages **a,** $\langle\theta_{m,l}\rangle$ and **b,** $\langle\varphi_{m,l}\rangle$ for the photonic circuits of 100 $U_n$ realizations with *n* = 64. We set the upper bound of the colormap in **a** to be $0.3\pi$ for better visibility.